\documentclass[12pt]{article}

\usepackage{epstopdf}
\usepackage[caption=false]{subfig}
\usepackage{graphics,graphicx,fullpage}

\usepackage{amsmath}
\usepackage{amssymb}
\usepackage[sans]{dsfont}
\usepackage{xcolor,color}
\usepackage{multirow, makecell} 

\usepackage[numbers,sort&compress]{natbib}
\bibpunct[, ]{[}{]}{,}{n}{,}{,}

\def\ga{\mbox{Ga}}
\def\no{\mbox{N}}
\def\un{\mbox{U}}
\def\data{\mbox{data}}

\newcommand{\Prob}{\mathbb{P}}

\newcommand{\Indicator}{\operatorname{\mbox{\large $\mathds{1}$}}}

\begin{document}

\baselineskip=24pt







\title{\bf Dealing with missing data under stratified sampling designs where strata are study domains}
\author{{\sc Carlos Rodr\'iguez, Luis Nieto-Barajas \& Carlos P\'erez-P\'erez} \\[2mm]
{\sl IIMAS-UNAM, ITAM \& GNP} \\[2mm]
{\small {\tt carloserwin@sigma.iimas.unam.mx, lnieto@itam.mx \& csampez@gmail.com}} \\}
\date{}
\maketitle

\maketitle

\begin{abstract}
{A quick count seeks to estimate the voting trends of an election and communicate them to the population on the evening of the same day of the election. In quick counts, the sampling is based on a stratified design of polling stations. Voting information is gathered gradually, often with no guarantee of obtaining the complete sample or even information in all the strata. However, accurate interval estimates with partial information must be obtained. Furthermore, this becomes more challenging if the strata are additionally study domains. To produce partial estimates, two strategies are proposed: 1) A Bayesian model using a dynamic post-stratification strategy and a single imputation process defined after a thorough analysis of historic voting information. Additionally, a credibility level correction is included to solve the underestimation of the variance; 2) a frequentist alternative that combines standard multiple imputation ideas with classic sampling techniques to obtain estimates under a missing information framework. Both solutions are illustrated and compared using information from the 2021 quick count. The aim was to estimate the composition of the Chamber of Deputies in Mexico.
}
\end{abstract}

\vspace{0.2in} \noindent {\sl Keywords}: Clustering; finite population sampling; missing data; multiple imputation; quick count; post-stratification

\section{Introduction}

\subsection{{Motivation}}
The 2021 Mexican quick count aimed to estimate the final composition of the Chamber of Deputies (COD) in the federal election. 500 deputies integrate the COD in Mexico: 300 are elected by simple majority in each of the 300 federal districts, and 200 are assigned by a proportional representation rule. Therefore, inference is required at a district level, as well as a population level. The natural sample design is a stratified random sampling by federal districts, with the particular aspect that strata are also study domains. Ultimately, the quick count produces interval estimates of the number of deputies (seats) obtained by each political force. This task is non-trivial because several rules have to be followed to transform votes into seats, these are described in \cite{CPEUM} and \cite{LEGIPE}. 

The primary challenge to overcome is the production of estimates under a missing information scenario, because of the gradual arrival of the sample of polling stations. On the election day, polling places close approximately at 18 hours. Hence, information from the sample of polling stations starts arriving gradually around 19 hours as it becomes available. These incomplete samples are updated every 5 minutes with new polling stations. The aim of the quick count is to produce partial estimates until the final estimation is obtained around 22 hours (or as soon as possible). Thus, it is impossible to wait until the full sample arrives because that might never happen. In addition, at the time of the final estimation,there could be strata with no information.

Missing data is a ubiquitous problem that complicates the statistical analysis of data collected in almost every discipline. When the information collected from a study is incomplete, there are important implications for their analysis. It should be clear that when a portion of the data is missing, there is an inevitably loss of information which implies a reduction in the parameters' estimation precision. In certain circumstances, missing data can introduce bias and thereby lead to misleading inferences about the parameters of interest. The validity of any method of analysis will thus require certain assumptions, often referred as the missing data mechanism. For an introduction on missing data ideas and implications see \cite{little} and \cite{hmdm2014}. 

In general, there are three missing data scenarios \cite{little}: Missing completely at random (if the events that lead to data being missing are independent both of observable variables and unobservable parameters of interest, and they occur solely at random), missing at random (when the missing values are not random, but these can be completely accounted for by variables where there is complete information), and missing not at random (there is a relationship between the propensity of a value to be missing and its values). It is well known, by all the members of the quick count committee, that there is bias in the arrival of the information. This is due to factors like the size of the nominal list\footnote{The nominal list is the register of citizens entitled to vote.} or the location (rural-urban) of the polling stations. These scenarios fall under a missing at random mechanism. However, some additional causes induce bias and enter under a missing not at random setting. For example, disagreements concerning voting results between voters and/or between voting officials and representatives of the political parties, etc.

\subsection{{Objectives}}

The main objective of this work is to propose two strategies to produce partial estimates, under missing data, in stratified random sampling where strata are also study domains. 

The first strategy is based on a frequentist estimation procedure and relies on multiple imputation ideas to deal with the missing information problem. {See \cite{rubin} for the original idea of multiple imputations and \cite{murray} for a review}. First, $m$ complete stratified samples are obtained, second sampling techniques are used to produce variance and point estimates for these $m$ stratified samples. Finally, these estimates are pooled into a single interval estimate. A Bootstrap alternative \cite{boot1}, along with a correction to obtain unbiased estimates under stratified random sampling is used. {See \cite{sitter1} and \cite{sitter2} for the solution used in our implementation and \cite{bootfin} for a recent discussion on the use of Bootstrap ideas under complex survey sampling}. The second strategy is based on a Bayesian estimation procedure and uses a dynamic post-stratification to impute the sufficient statistics of the Bayesian model. The post-stratification is defined by a thorough analysis of historic information. A further credibility level correction is suggested, to solve the underestimation of the variance when a single imputation is used. Post-stratification is a technique for adjusting a non-representative sample after the sample has been selected so it tackles, among many issues, missing information \cite{glasgow}. See also \cite{anza:gonz:ortiz} for a recent application of post-stratification using multilevel regression in the context of the quick counts for Mexican governorship elections.


A minor objective is to show how to determine the sample size and estimation errors in the stratified sampling design for the 2021 Mexican quick count.

\section{The Chamber of Deputies}\label{sec:COD}

The COD in Mexico is conformed by 500 deputies, 300 of them by the principle of simple (or relative) majority in each of the 300 federal districts\footnote{Geographically Mexico is divided into 300 federal districts. These districts are delimited to distribute the population of the country in a balanced manner so that each elected deputation represents a similar number of inhabitants.} and 200 by proportional representation.


For the principle of simple majority, it is important to consider the coalitions between different political forces. If there is a coalition in a particular district, the sum of the votes obtained by all valid ways of voting for the coalition will be used in favor of the coalition candidate. In case the coalition wins, the parties agree for which one the seat in the COD will be assigned. To determine the weight of each party at the national level it is important to distribute the coalition votes among the parties that made up the coalition. In this case, the votes for valid combinations that include two or more parties are equally divided among each party in the combination. However, individual votes for each party are not shared. 

In the case of proportional representation, two issues are important to mention. The first is that only political parties that accumulate more than 3\% of the votes at a national level can enter the distribution of deputies by this principle\footnote{Independent candidates, null plus unregistered votes and abstentions are not considered for proportional representation.}. Second, there are two upper bounds for the total number of deputies obtained by each party: 1) no political party can have more than 300 deputies, and 2) no political party can have a proportion of deputies that exceeds 8\% of its proportion of votes at a national level. These two bounds imply an iterative process to assign the deputies by proportional representation, and this is outlined below. 

Let $x_{h,j,r}$ denote the votes in favour of the political force $j$ in stratum $h$ and polling station $r$. Then, the maximum number of deputies for the political force $j$ is given by $M_j=\min\left\{300, \left\lfloor{500\left(\eta_j + 0.08\right)}\right\rfloor \right\}$, 
where $\eta_j = x_{.j.}/x_{...}$ and with $x_{.j.}$ the total votes for party $j$ at a national level, and $x_{...}$ is the total number of votes for all political forces that have more than 3\% of the votes at a national level. 

In the first distribution by proportional representation, 200 deputies are distributed considering $200 \eta_j$. Let us assume that after receiving the deputies by proportional representation, the total number of deputies obtained by, say $j=1$, considering both principles, exceeds $M_1$. Then, an adjustment is made to the number of deputies by proportional representation such that the total number of deputies is exactly $M_1$ and force $j=1$ is removed from the distribution. If the number of deputies by proportional representation is given by $R_1$, in the next step of the iteration $200 - R_1$ deputies are distributed considering $(200-R_1)(x_{.j.}/(x_{...} - x_{.1.}))$ among the remaining political parties. Hence, this process is repeated until all the political forces meet both bounds. A final concern is the rounding problem in integers quantities, e.g. $200 \eta_j$, and this is solved by the largest residual method.

This is a very particular algorithm and is carefully described in \cite{CPEUM} and \cite{LEGIPE}. However, the inputs that it requires are the total number of votes $x_{h,j.}$ for the political force $j$ in all the federal districts $h$. With these totals at the federal district level, it is trivial to obtain $x_{.j.}$ and $x_{...}$ as described in the paragraph above. Hence, the attention is focused on the estimation of these totals and the use of the algorithm to transform votes into seats granted for the COD.

\section{Estimation methods under complete samples}\label{sec:estimation}
{This section describes the strategies to estimate the conformation of the COD. We work under a stratified sample setting assuming the complete sample on each stratum is available, and the strata are study domains. First, the notation for the usual stratified sampling is introduced and then the frequentists and Bayesian estimating approaches are outlined.}

\subsection{Notation}
The sample design is based on a stratified random sampling of polling stations by federal district with 300 strata, where strata are also study domains. 
It is important to stress that when a polling station is selected for the quick count, the votes for all political forces including political parties, independent candidates plus null, unregistered votes and abstentions are collected. So, in principle, complete results on each polling station in the sample are available to make inference. 

Before proceeding we introduce the basic concepts, as well as the notation used by the two estimation strategies. In stratified random sampling the finite population, consisting of $N$ units is partitioned into $L$ non-overlapping strata of $N_h$ units, for $h=1,\ldots,L$; clearly $N=\sum_{h=1}^L N_h$. To estimate the composition of the COD, $N$ is the total number of polling stations installed at a national level, $N_h$ are the polling stations installed in the federal district $h$, with $L=300$ federal districts. A simple random sample without replacement (SRSWOR) of polling stations is taken independently from each stratum. The sample sizes within each stratum are denoted by $n_h$, for $h=1,\ldots,L$; thus the total sample size is $n=\sum_{h=1}^L n_h$. Finally, $x_{h,j,r}$ denotes the votes in favour of the political force $j$ in the stratum $h$ and polling station $r$, where $ j = 1, \ldots, J $ and $J$ is the total number of political forces (which includes political parties and independent candidates) plus two extra categories: null plus unregistered votes and abstentions. The population parameter of interest $\psi= \psi(\boldsymbol{S}) $, where $\boldsymbol{S}= \{x_{h,j,r}| h = 1, \ldots, L;  j = 1, \ldots, J;  r = 1, \ldots, N_h \}$, is usually estimated by  $\widehat{\psi}= \widehat{\psi}(\boldsymbol{s}) $, where $\boldsymbol{s}= \{x_{h,j,r}| h = 1, \ldots, L;  j = 1, \ldots, J;  r = 1, \ldots, n_h \}$.

\subsection{Frequentist method}\label{boot0}

The stratified sampling design is defined via classic sampling techniques (non-parametric), and thus to obtain estimates of the total votes for each political force $j$ is trivial, e.g. \cite{kalton}. However, we need to follow classic re-sampling techniques (e.g. \cite{boot2}) to obtain interval estimates for the conformation of the COD. {It is well known that when the usual Bootstrap is used to estimate totals under a stratified sampling approach the variance estimators are biased, e.g. \cite{sitter2}}. Thus, the ideas of \cite{sitter1} were used to obtain unbiased estimates for the total number of votes. For completion, an outline of this strategy is provided.

Let $f_h= n_h/N_h$ be the probability of selecting a polling station at a stratum $h$, then define $k_h=1/f_h$ and $m_h= f_h n_h$, for $h=1, \ldots, L$ and assume $k_h$ and $m_h$ for all $h$ are integers. The following steps are iterated $B$ times:
\begin{enumerate}
	\item Set $h=1$.
	\item {Select a SRSWOR of $m_h$ polling stations from the $n_h$ selected in district $h$ to get $x^*_{hjr}$ for $j = 1, \ldots, J$ and $r=1, \ldots, m_h$}.	
	\item Repeat the last step $k_h$ times and thus obtaining a sample of $n_h=k_h m_h$ polling stations.
	\item The Bootstrap estimate for the total votes in stratum $h$ for the political force $j$ is thus given by
	$\widehat{x}^*_{hj.}= \displaystyle\frac{N_h}{n_h} \displaystyle\sum_{r=1}^{n_h} x^*_{hjr},\ \mbox{for} \ \ j = 1, \ldots, J.$	
	\item If there are coalitions, thus considering the coalition agreements;
	\begin{enumerate}
		\item Estimate the political force that wins the seat in the Chamber. 
		\item Votes for valid combinations of political parties in the coalition, that include two or more parties, are divided between each party in the combination. The remainder of the division is assigned to the political party with the most votes.
	\end{enumerate}
	\item Set $h = h + 1$ and return to Step 2 while $h \leq L$.
	\item Estimate the total votes in favor of each political force at a national level. 
	\item Apply the algorithm to obtain the conformation of the COD.  {This algorithm further requires estimates of 
	\begin{itemize}
	\setlength\itemsep{0.2em}
		\item The total number of votes obtained by each political force at a national level, i.e
		$\widehat{\nu}^*_j= \displaystyle\sum_{h=1}^{300} \widehat{x}^*_{hj.},\ \ \mbox{for}\ \ j=1, \ldots, J.$
		\item The proportion of valid votes in favor of each political force at a national level, and these are obtained via $
	\widehat{\lambda}^*_j = \displaystyle \frac{\widehat{\nu}^*_j}{\sum_{j=1}^{J-2} \widehat{\nu}^*_j},\ \ \mbox{for}\ \ j=1, \ldots, J-2,$
	       where null and unregistered votes are ignored. 
		\item The proportion of valid votes for each political party after excluding parties that do not reach $3\%$ in valid voting (independent candidates are ignored), this is given by 
           $ \widehat{\eta}^*_j = \displaystyle \frac{\widehat{\nu}^*_j \Indicator(\lambda^*_j > 0.03)}{\sum_{j = 1} ^{J-3} \widehat{\nu}^*_j \Indicator(\lambda^*_j > 0.03)},\ \ \mbox{for}\ \ j=1, \ldots, J-3,$
	\end{itemize}
	where $\Indicator(A)$ denotes the indicator function of set $A$.}
\end{enumerate} 
{The algorithm produces $B$ conformations of the COD. Confidence intervals for the number of seats for each political force are obtained via the percentile method, see e.g. \cite{wasserman}.}

If for some strata $k_h$ and $m_h$ are not integers there is a randomization process to round these to an integer and the algorithm is applied exactly as above. This randomization process ensures the estimates are unbiased, see \cite{sitter1}.  {Steps 1 to 4 are described in \cite{sitter2} (2.2. The Mirror-Match Method)}.

\subsection{Bayesian method}\label{baycomp}

Although the rules to obtain the conformation of the COD are based on the total votes per party, it is possible to represent the criteria in terms of the proportions of votes, that we introduce with the following notation. Let $l_{h,r}$ be the size of the nominal list, that is, the size of the list of all citizens who have the right to vote, in polling station $r$ of stratum $h$. Let $\theta_{h,j}$ be the proportion of votes, relative to the size nominal list $l_{h,r}$, in favour of the political force $j$ in the stratum $h$, with $ j = 1, \ldots, J $ and $h= 1, \ldots, L$. Let $l_h=\sum_{r=1}^{N_h} l_{h,r}$ the size of the nominal list of voters in district $ h $, for $ h = 1, \ldots, L $, and $ l = \sum_{h= 1}^L l_h $ the size of the national nominal list. Then $\theta_j = \sum_{h = 1}^L \left(l_h/l\right) \theta_{h, j} $ is the proportion of votes, relative to the size nominal list, in favour of political force $j$ at national level.

For the estimation of $\theta_{h, j}$, we use a slight modification of the model proposed by \cite{mendoza:nieto}. This model assumes the votes in favour of the political force $j$ in the stratum $h$ in the polling station $r$ follows the distribution
\begin{equation}
\nonumber
X_{h,j,r} | \theta_{h,j}, \tau_{h,j} \sim \no\left (x_{h,j,r} \bigg|  l_{h,r} \theta_{h,j}, \frac{\tau_{h,j}}{l_{h,r}} \right), 
\end{equation}
for $r=1,\ldots,n_h$, where $\tau_{h,j}$ is the precision assumed constant within the stratum and independent of $\theta_{h,j}$. Even more, \(X_{h,j,r}\) is assumed independent of \(X_{h,j',r}\), for all \(j \neq j' \).

{The original prior distribution for the unknown model parameters $(\theta_{h, j},\tau_{h, j})$, considered in \cite{mendoza:nieto}, assumes independence and is of the form $\theta_{h,j}\sim\un(0,1)$ and $\tau_{h,j}\propto\tau_{h,j} ^{-1} \Indicator(\tau_{h,j}> 0)$. However, this prior only produces a proper posterior if sample size $n_h\geq 2$, wasting valuable information when we only have a single polling station. To overcome this problem we now propose to use a proper but vague prior of the form $\tau_{i,j}\sim\ga(0.5,0.05)$, which has a large variance of 200.}

{Thus, using Bayes' Theorem, the posterior distribution turns out to be proportional to the product of a truncated normal for \(\theta_{h,j} \) conditional on \(\tau_{h,j} \) and a gamma distribution for \(\tau_{h,j} \). Specifically we have
\begin{align}
\nonumber
p(\theta_{h,j},\tau_{h,j}\mid\data)&\propto\no\left( \theta_{h,j}\left| \frac{\sum_{r=1}^{n_h}x_{h,j,r}}{\sum_{r=1}^{n_h}l_{h,r}}\,,\,\tau_{h,j}\sum_{r=1}^{n_h}l_{h,r} \right.\right) \Indicator(0<\theta_{h,j}<1) \\
\label{eq:posterior}
&\times\ga\left( \tau_{h,j}\left| \frac{n_h}{2},\,\frac{1}{2}\left\{ \frac{1}{10}+\sum_{r=1}^{n_h}\frac{x_{h,j,r}^2}{l_{h,r}}-\frac{\left(\sum_{r=1}^{n_h}x_{h,j,r}\right)^2}{\sum_{r=1}^{n_h}l_{h,r}} \right\} \right.\right).
\end{align}
}
To determine the winning deputies by \textit{simple majority} we define
\begin{equation}
\nonumber
\Indicator(\theta_{h,j} > \theta_{h,k}),\ \forall\ k \neq j, k = 1, \ldots, J-2,
\end{equation}
an indicator that tells us whether or not political force $j $ won in district $h$. 

To determine the winning deputies by \textit{proportional representation}, we first define $ \lambda_j = \theta_j / \sum_{k = 1} ^{J-2} \theta_k $, the proportion of valid votes (nulls plus no registered and abstentions are excluded), and $ \eta_j = \theta_j / \sum_{k = 1} ^{J-3} \theta_k  \Indicator(\lambda_j > 0.03)$, as the proportion after excluding parties that do not reach $3\%$ in valid voting, independents, null and unregistered.

{Posterior point and interval estimates for $ \theta_{h,j} $ are approximated via simulations from the posterior distribution. This is required due to the truncation. For details see \cite{mendoza:nieto}. For each posterior draw, a conformation of the COD is obtained. Highest posterior density intervals for the number of seats for each political force are finally obtained using the 2.5\% and 97.5\% quantiles.}

\section{Incomplete Sample Strategies}\label{sec:incomplete}
{This part describes the strategies to obtain estimates under an incomplete sample setting. Both proposals use imputation strategies to estimate employing the methods for complete samples. As a motivation, the complete sample sizes and the sample sizes received in several quick counts are presented first.}

\subsection{Motivation: previous quick counts}

There has not been any quick count that has received the complete sample at the appropriate time. In Table \ref{tab3}, the total sample size and the sample size used to calculate the last estimation for six quick counts are displayed, 
three for the COD, and three presidential. Note that values for the 2021 election are also included, the total sample size was $6,345$ polling stations. In the next section results from a simulation study are shown to justify this sample size. Table \ref{tab3} indicates the worse scenario was in 2018 (COD election) where only 67.5\% of the sample was used to produce final estimates, and the best was in 2006 (presidential election) where 95.1\% of the sample was used to produce estimates. 
\begin{table}[!htbp]
\centering
\begin{tabular}{ccccccc}
  Election& \multicolumn{3}{c}{COD} & \multicolumn{3}{c}{Presidential}\\ 
  \hline
  Year                    & 2003    & 2015        & 2021   & 2006   & 2012    & 2018    \\ \hline
  Total polling stations ($N$)  &   121,284  & 149,726   & 163,666  &   130,788   & 143,437  &  156,840  \\
  Total sample size  ($n$)            & 7,236   & 9,450       & 6,345  & 7,636  & 7,597   &  7,787  \\ 
  Received sample size       &  6,743  & 7,016       & 5,040  & 7,263  & 6,260  & 5,254   \\
  \%  of SS received        &   93.1   &  74.2        &   79.4  &  95.1   &   82.4  &   67.5 \\   
  Time of final estimate        & 22:00   & 22:15       &   22:35  & 22:15  & 22:30  &  22:30   \\       
  \hline
\end{tabular}
\caption{Total polling stations, quick count sample sizes (total, received and percentage) and time of final estimate.}\label{tab3}
\end{table}

We are clearly facing a missing data problem. One simple solution could be to define a larger sample size (sample augmentation), however there are two important factors to consider: The pressure the quick count exerts on the field people must be minimized, and also the bias in the arrival of the information due to factors of the polling stations, such as the size of the nominal list, the location in a rural or urban area, causes under a missing not at random setting. Therefore, larger sample sizes would increase bias and would not solve the missing data problem.

The total number of polling stations is growing year after year as it depends on the population size. However, the sample size must be kept small enough. 
For the quick count to be useful, final estimates have to be produced around the 22 hours (or earlier if possible) on the election day. Then, it is essential to produce estimates with incomplete samples. Here two solutions are proposed. 

\subsection{Frequentist alternative using multiple imputation}

One direct way to deal with the missing data problem is via multiple imputation (MI), which is a three-stage approach. First, missing values are imputed $m$ times using a statistical model based on the available data. Thus, effectively generating $m$ complete datasets. Second, inference is performed on each of the $m$ complete data sets. {On this step the Bootstrap algorithm described in Section \ref{boot0} is used on each of the $m$ complete data sets.} Third, the $m$ estimates are consolidated/pooled to obtain a single estimate. 

It is easy to understand how several  {multiple univariate imputation strategies} work by considering these as something similar to making predictions under a regression context. A set of predictor variables $\boldsymbol{X}$ are needed to describe a dependent variable $Y$, and the aim is to predict the values of the dependent variable that are missing. Observations with full information are used to fit the model, and predictions for the missing observations are the imputed values. However, since $m$ different predictions/imputations for the same missing observation are required, a randomization process is included. In our case the predictive mean matching (\texttt{pmm}) algorithm \cite{little} was used. For completion, an outline of the procedure is provided. First, fit a linear regression model via Bayesian ideas. Second, draw randomly from the posterior predictive distribution of $\boldsymbol{\beta}$ and produce a new set of regression coefficients $\boldsymbol{\beta}^*$. Third, calculate predicted values for observed and missing $Y$, and for each case where $Y$ is missing, find the closest predicted values among cases where $Y$ is observed. The \texttt{pmm} algorithm selects the closest observed values (typically with 3, 5 or 10 members) to the missing value $y_i$. Fourth, draw randomly one of these close cases and impute the missing value $y_i$ with the observed value of this close case. Finally, under an MI setting these steps are repeated $m$ times.

{To deal with the multivariate case, the MI strategy can be extended as follows. First, to start the algorithm the mean of the non-missing observations per variable is imputed on each missing value. Second, mean imputations for the first variable are put back as missing and the univariate imputation strategy is used to impute all the missing values of this variable. The remaining variables are used as predictor variables. Third, mean imputations for the second variable are put back as missing and the univariate imputation strategy is used to impute this variable. This process continues until all the missing values for all the variables have been imputed. This is one iteration of the imputation process. We can define the number of iterations to perform, the variables to use as predictors, and even the order in which each variable is used. This algorithm has been implemented in R via the \texttt{mice} library \cite{buuren}. In our case, only the major political parties MORENA, PAN, PRI, and PRD, along with the nominal list were used as predictor variables. The variables were imputed from left to right, this means older political parties were imputed first. Finally, the number of iterations was set to 5. }

MI can be used in cases where the data are missing completely at random, missing at random, and even when the data are missing not at random, see e.g. \cite{buuren0}. Thus, this appears as a reliable solution for the problem at hand. The frequentist strategy is as follows:
\begin{enumerate}
  \item Via MI generate $m=15$ complete data sets of $n$ polling stations. To perform MI we use the \texttt{mice} library of R see \cite{buuren0} and \cite{buuren}.  {As mentioned before the \texttt{pmm} algorithm was employed to generate the imputed datasets, where only the major political parties and the nominal list were used as predictor variables}. These predictor variables were used to impute all the variables with missing values.
   \item {Run the Bootstrap strategy described in Section \ref{boot0} setting $B=300$ to obtain estimates for the COD with each complete dataset. Here parallel processing was used to reduce the computational time, see the libraries \texttt{foreach} and \texttt{doParallel} of R, see \cite{foreach} and \cite{do}. }
 \item Consolidate these $m=15$ estimates for the conformation of the COD. This was done following the pooling formulas 3.1.2 to 3.2.10 in \cite{rubin}.
\end{enumerate} 

\subsection{Bayesian strategy via dynamic post-stratification}

When estimating totals or simple percentages at the population level under a stratified sampling design with no information available in all the strata, it is customary to follow a post-stratification approach and obtain estimates via this alternative. However, if the original strata are also study domains we still have to make inference for them. To solve this problem, an imputation step based on a dynamic post-stratification is used for the missing strata, and inference is made using the complete sample Bayesian method described in Section \ref{baycomp}. {Because with incomplete samples coverage of interval estimates is a lot lower than the nominal credible level set, a further adjustment is made to the credible level according to the percentage of the sample received}. The proposal relies on imputing the sufficient statistics, which according to the posterior distribution \eqref{eq:posterior} for $(\theta_{h,j},\tau_{h,j})$, are $T_{h,j}=\sum_{r=1}^{n_h}X_{h,j,r}$ and $U_{h,j}=\sum_{r=1}^{n_h}\frac{X_{h,j,r}^2}{l_{h,r}}$. Additionally, define $v_{h,j}=\sum_{r=1}^{n_h}l_{h,r}$. The proposal is outlined as follows:
\begin{enumerate}
\item Define a sequence of clusterings, starting with the trivial clustering $P_1$, which contains all the strata in a single group, up to the sample design stratification $P_L$ where each stratum belongs to a different group. {These clusterings are predefined using historic data. In particular, voting results of the 2018 federal election were used to characterize each stratum and define the post-stratifications: a hierarchical clustering technique with complete linkage along with the standardized Euclidean distance was used.} To be used in the 2021 quick count, seven post-stratifications where pre-defined: $P_k$ for $ k \in \{1, 10, 20, 50, 100, 200, 300 \}$. 
\item For strata with complete or partial information, inference is made using the available data. For strata with no information, say $h$, let $P_k^{(h)}$ the specific group of clustering $P_k$ that contains stratum $h$. Determine $k^*$, the largest $k$ such that $P_{k}^{(h)}$ has sample information available. Considering the poststratification group $P_{k^*}^{(h)}$, averages of the sufficient statistics over the strata with information are obtained, and impute these on stratum $h$ with further assuming that the statistic is the result of observing a single polling station with maximum nominal list of 750 voters. In notation, the imputed statistics for stratum $h$ are
$$T_{h,j}=l_0\frac{\sum_{h'}^{m_{h,k^*}}T_{h',j}}{\sum_{h'}^{m_{h,k^*}}v_{h',j}} \quad\mbox{and}\quad U_{h,j}=l_0\frac{\sum_{h'}^{m_{h,k^*}}U_{h',j}}{\sum_{h'}^{m_{h,k^*}}v_{h',j}},$$
where $m_{h,k^*}$ is the number of strata with sample information in $P_{k^*}^{(h)}$, $l_0=750$ and set $n_h=1$ and $l_{h,r}=l_0$ for inference purposes.
\item Run the algorithm outlined in Section \ref{baycomp}.
\item {To match the empirical coverage with the desired credible level, in the presence of incomplete samples, interval estimates are obtained by adjusting the credible level as described in Table \ref{tab:credible}. These corrections were based on fitting an accelerated failure time model to the arrival time of data from the 2018 federal election and using the features and votes in the polling stations as explanatory variables. Later, a simulation study was implemented by creating incomplete samples according to the predicted arrival time and computing the coverage for different credible nominal levels. Therefore, as suggested by Table \ref{tab:credible}, if the percentage of the sample received is less than $60\%$, estimates at a $99\%$ credible level are produced. The credible level reduces gradually until more than $90\%$ of the sample is reached. Thus producing estimates at a $95\%$ credible level, which is the target level set for the 2021 quick count.} 
\end{enumerate}

\begin{table}[!htbp]
\begin{center}
\begin{tabular}{lccccc}
\hline\\[-2mm]
Percentage & $[0,60)$ & $[60,70)$ & $[70,80)$ & $[80,90)$ & $[90,100]$ \\
Credibility & $99$ & $98$ & $97$ & $96$ & $95$ \\[1mm]
\hline
\end{tabular}
\caption{Credibility level adjustment according to percentage of sample received.} \label{tab:credible}
\end{center}
\end{table}

\section{Sampling design for the 2021 quick count}\label{sec:design}
{Here the considerations taken to determine the total sample size, $n$, used to estimate the COD for the 2021 quick count are described. A stratified sample by federal district represents a reasonable choice and the federal districts are also study domains.} 

\subsection{Estimation errors}
The stratified sampling design by federal district requires to determine a winner in each of the strata, therefore each stratum is equally important. Thus it was decided to set $n_h = n/300$, for $h=1,\ldots,300$. In this case, only the overall sample size $n$ needs to be defined. This also helps to deal with the non-response, avoiding federal districts with very small sample sizes. 

The inference problem in the quick count is to obtain an estimator $\widehat{ND}_j$ of the true number of deputies $ND_j$, for each party $j=1,\ldots,J-2$. Then, the estimation error for each party is simply $|ND_j-\widehat{ND}_j|$. To summarize the margin of error of our sample design we use two measures: the average and the maximum estimation errors of all parties. For each of them, we want to determine upper bounds $\epsilon_1$ and $\epsilon_2$ such that
\begin{equation}
\nonumber
\Prob\left(\frac{1}{J-2} \sum_{j=1}^{J-2} |ND_j - \widehat{ND}_j|  \leq \epsilon_1 \right)  = 0{.}95 \ \ \mbox{and}\ \
\Prob\left(\max_{j=1,\ldots, J-2} |ND_j - \widehat{ND}_j|  \leq \epsilon_2 \right)= 0{.}95. 
\end{equation} 
Simulation techniques based on repeated sampling and using the classic and Bayesian estimation methods described above were used to approximate $\epsilon_1$ and $\epsilon_2$ for different sample sizes. 

Table \ref{margin} displays estimates of $\epsilon_1$ and $\epsilon_2$ for various sample sizes. These were calculated using the voting results for the COD elections of 2012, 2015, and 2018. It is important to note that in each election there were different numbers of installed polling stations and also different numbers of political parties. In $2012$ there were seven political parties, ten in $2015$, and nine in $2018$. Table \ref{margin} also shows that the average estimation errors change drastically from 2015 to 2018. Instead, the maximum is more stable across the different elections. Thus, considering the maximum it is clear that there is no big improvement in precision between setting $n_h=15$ or $n_h=30$. However, it was essential to obtain estimates around the 22 hours of the 6$th$ of June, 2021. Hence, with these considerations, it was decided to select a sample size of $n_h=20$ polling stations per stratum. This implies an overall sample size of $n = 6,000$ polling stations.

\begin{table}[ht]
\centering
\begin{tabular}{|c|cc|cc|cc|}
\hline
  \multirow{2}{*}{$n$ ($n_h$)}    & \multicolumn{2}{c|}{2012} & \multicolumn{2}{c|}{2015} & \multicolumn{2}{c|}{2018} \\
  &  $\hat{\epsilon}_1$ & $\hat{\epsilon}_2$  &  $\hat{\epsilon}_1$ & $\hat{\epsilon}_2$ &  $\hat{\epsilon}_1$ & $\hat{\epsilon}_2$ \\ 
 \hline
1,500 (5) & 2.9 & 9    & 2.8 & 10    & 0.7 & 9\\ 
3,000 (10) & 2.3 & 7  & 2.4 & 9    & 0.5 & 7\\ 
4,500 (15) & 2.0 & 6  & 2.3 & 9    & 0.4 & 6\\ 
6,000 (20) & 2.0 & 6  & 2.2 & 9    & 0.4 & 6\\ 
7,500 (25) & 1.7 & 5  & 2.0 & 8    & 0.3 & 5\\ 
9,000 (30) & 1.7 & 5  & 2.0 & 8    & 0.3 & 5\\ 
  \hline
\end{tabular}
\caption{Upper bounds for average and maximum estimation errors at 95\% confidence level for different samples sizes. Reported numbers correspond to the COD elections of 2012, 2015 and 2018.}\label{margin}
\end{table}

\subsection{Sample augmentation}\label{sd2}

The quick count needs to produce interval estimates around the 22 hours of the election day. However, in Mexico there are six states with different time zones: Baja California (8 districts) and Sonora (7 districts) are two hours behind the center time; Chihuahua (9 districts), Baja California Sur (2 districts), Nayarit (3 districts) and Sinaloa (7 districts) are one hour behind. Thus, the sample size in the strata of the states two hour behind was increased in 10 polling stations, and in the strata of the states with one hour behind additional 5 polling stations were added. Also, historically the response in Guerrero (9 districts) has been very low, thus an additional 10 polling stations were added to each of its strata. With these considerations, the total sample size to estimate the conformation of the COD in the 2021 election was $n = 6,345$ polling stations.

\section{Election Day}\label{sec:election}
{This section describes the arrival of information and estimates for the 2021 quick count. The total sample size was of $n = 6,345$ polling stations. However, the sample is gathered gradually. Thus, our proposals to estimate the conformation of the COD under incomplete samples were used several times during the evening and night of the 6$^{th}$ of June 2021. These estimates are used to build a time series, allowing us to assess the performance of each proposal.} 

\subsection{Information arrival}
\begin{figure}[!htbp]
\begin{center}
	\includegraphics[scale=0.54]{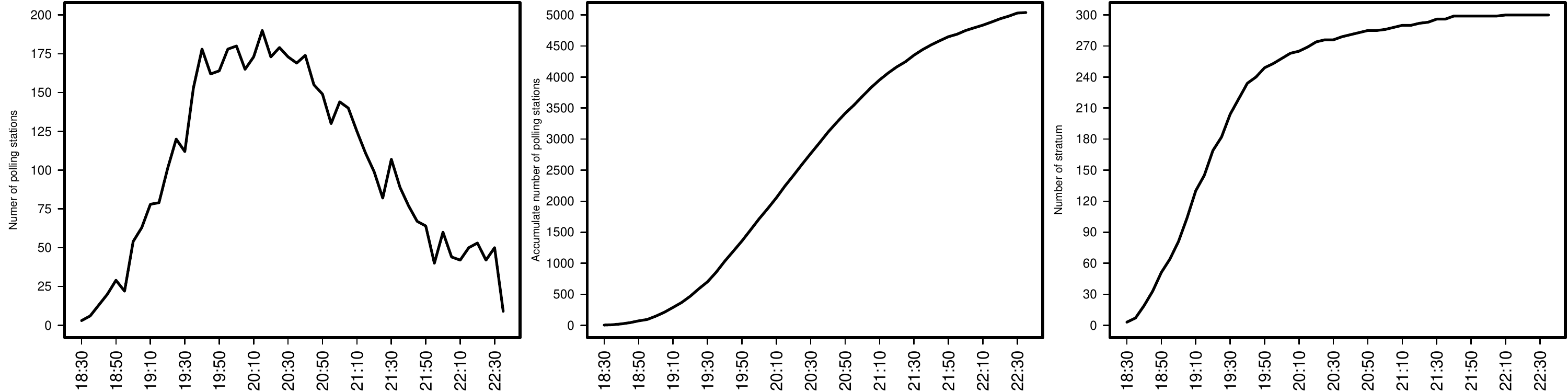}
\end{center}
\caption{Arrival of information summaries (time series): number of new polling stations received on each update (left), cumulative number of polling stations (center) and number of strata with sample data received.} \label{fig1}
\end{figure}

On the 6$^{th}$ of June 2021, voting information of the first polling stations from the sample for the quick count was received at 18:30 hours: 3 polling stations of 3 different stratum. From then, every data update contained an average of 100 new polling stations more than the previous update. The peak of information arrival was attained between 19:30 and 20:30 hours. Approximately, 175 new polling stations were received on every update within this time interval. At this point, around 3,000 polling stations had been received overall, and observed that 250 stratum had at least one polling station. Then, the rhythm of information arrival decreased and the main concern was to receive data in each of the 300 strata, with at least two polling stations in each. Figure \ref{fig1} shows summaries of the number of polling stations received on each update.

\subsection{Estimations}
\begin{figure}[!htbp]
\begin{center}
	\includegraphics[scale=0.76]{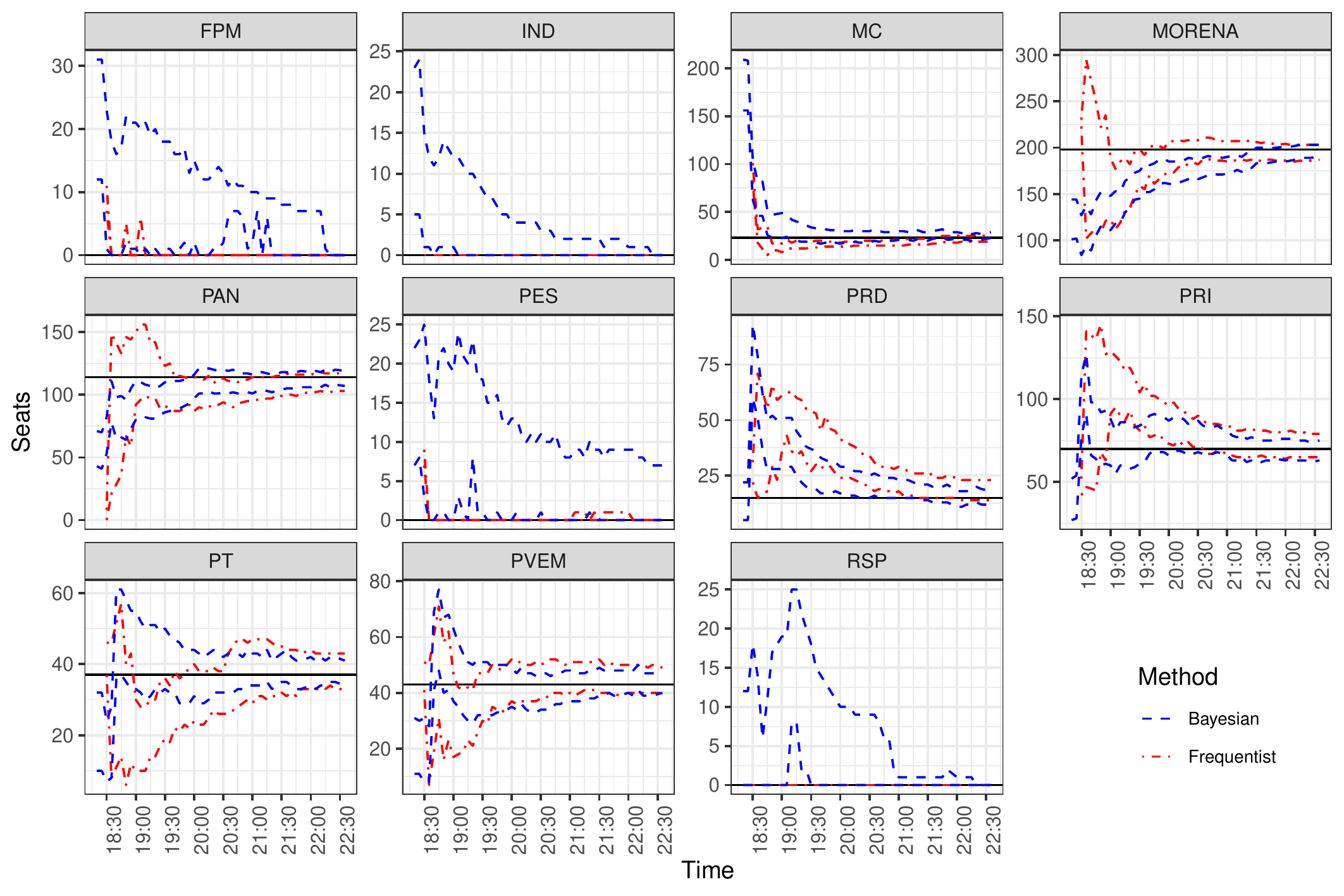}
\end{center}
\caption{Bayesian (dashed lines) and frequentist (dashed-dotted lines) estimations for the conformation of the COD for each political force in the 2021 quick count. Official final results are shown as solid horizontal lines.} 
\label{fig2}
\end{figure}

Estimates of the conformation of the COD were obtained considering every 5 minutes update of information using both methods presented in Section \ref{sec:incomplete}. These are shown in Figure \ref{fig2}.  Official final results have also been included as a solid thick line. Thus, it is straightforward to assess the performance of each strategy under an incomplete sample setting. 

At 18:30 hours the first estimation was obtained using only three polling stations. It is important to mention that the frequentist method based on MI did not impute values for all the valid voting combinations of the coalitions nor the independent candidates. This is easy to understand because there was no information available to predict values for these situations. Thus, in these cases, zeros were assigned. Therefore the first estimation was completely non-informative. This is reflected in the narrow confidence intervals reported. The same pattern continues until 18:45 hours, were 37 polling stations were received and confidence intervals started to behave as expected, i.e. reflecting the uncertainty due to the missing information in wide intervals. 

On the other hand, the Bayesian method reflected the uncertainty of the lack of information in wide intervals for the first data received, and gradually reduced the uncertainty by producing narrower intervals for the last updates of information.

In any case, both methods started to agree around 20:30 hours with complete agreement at 22:35 hours, when the final estimates were produced. Available data at this final time had 5,040 polling stations, 79.4\% of the complete sample of a total of 6,340 polling stations, and with information in all the design strata. The average number of polling stations available per stratum was 17, while the minimum and maximum were 3 and 28 respectively.



\section{Discussion}\label{sec:discussion}
 {We have proposed Bayesian and frequentist strategies to produce interval estimates, under stratified sampling designs, when there is no guarantee of obtaining the complete sample or even information in all the strata. Under a complete data scenario, both inference approaches produce very similar interval estimates. The differences we appreciate in Figure \ref{fig2} are due to the imputation criteria under a missing data scenario. The strategy used together with the frequentist approach imputes the votes at a polling station level, although the strategy proposed together with the bayesian approach imputes the sufficient statistic at a stratum level. As more data is available, both approaches coincide, 
therefore indicating we have solved the problem in a reasonably straightforward manner. Although, the search for better and more useful strategies is needed.}

Considering the Mexican Electoral National Institute has adopted the quick count as a permanent tool for every federal and local electoral process, it is important to review some of the challenges that the quick count of future electoral processes will have to face.

The first challenge concerns the bias in the arrival of the information and the fact that partial and final estimations are always computed under an incomplete sample setting. Since the quick count handles an extremely narrow margin of error, even with a minor relative bias, the estimated intervals could fail to contain the exact value. Naturally, it is recognized that with a 95\% confidence/credible interval, 5\% of the possible samples will lead to intervals that do not contain the true value. {However, in the presence of relative bias, the true coverage could be considerably lower than 95\%}.
 

The second challenge is to define a sample design that reduces the polling stations that every field people has to collect without sacrificing precision in the estimation. In practice, each field person has {4 polling stations assigned in average}, regardless of whether the station is in the sample or not. So, a same person could have to report data from more than one polling station. There have been some ideas for a two-stage sampling design. The obstacle over here is the estimation of the variance. This is because the frequentist formula for this sampling design requires at least two polling stations for every person selected.

Other problems that have been discussed are: To perform a verification/validation of the quality of the information at arrival; how to address situations when the estimated intervals of two or more political forces overlap with each other; or which is the most efficient strategy to consolidate two or more interval estimates for the same party, the union presumably is not the best alternative, etcetera.

\section*{Acknowledgements}

The authors are grateful to the Electoral Federal Registry of the Electoral National Institute for the ongoing support in organizing the quick count. In particular to Arturo Gonz\'alez Morales, former director of Statistics. 

\bibliographystyle{tfs}
\bibliography{References}

\end{document}